\documentclass[12pt,english]{revtex4}
\usepackage[T1]{fontenc}
\usepackage[latin1]{inputenc}

\makeatletter

\usepackage{babel}
\makeatother
\begin{document}

\title{Does the present data on $B_{s}-\bar{B}_{s}$ mixing rule out a large
enhancement in the branching ratio of $B_s \rightarrow\mu^{+}\mu^{-}$?}

\author{Ashutosh Kumar Alok and S. Uma Sankar}

\affiliation{Indian Institute of Technology, Bombay, Mumbai-400076, India}

\begin{abstract}

In this letter, we consider the constraints imposed by the recent
measurement of $B_s - \bar{B}_s$ mixing on the new physics contribution
to the rare decay $B_s \rightarrow \mu^+ \mu^-$. New physics in the
form vector and axial-vector couplings is already severely constrained by 
the data on $B \rightarrow (K,K^*) \mu^+ \mu^-$.
Here, we show that $B_s - \bar{B}_s$ mixing data, together with the
data on $K^0 - \bar{K}^0$ mixing and $K_L \rightarrow \mu^+ \mu^-$ 
decay rate, strongly constrain the scalar-pseudoscalar contribution
to $B_s \rightarrow \mu^+ \mu^-$. We conclude that new physics 
{\it can at best} lead to a factor of 2 increase in the branching
ratio of $B_s \rightarrow \mu^+ \mu^-$ compared to its Standard Model
expectation.
\end{abstract}
\maketitle

The flavour changing neutral interaction (FCNI) $b\rightarrow s \mu^{+}\mu^{-}$
serves as an important probe to test the Standard Model (SM) and its possible
extensions. This four fermion interaction gives rise to semi-leptonic
decays $B\rightarrow(K,K^{*})\mu^{+}\mu^{-}$ and also the purely leptonic
decay $B_{s}\rightarrow \mu^{+}\mu^{-}$. The semi-leptonic decays 
$B\rightarrow(K,K^{*})\mu^{+}\mu^{-}$ have
been observed experimentally \cite{babar-03,belle-03,babar-06} with
branching ratios close to their SM predictions 
\cite{ali-02,lunghi-02,kruger-01}. At present there is only an upper
limit, $1.0\times10^{-7}$ at 95\% C.L., 
on the branching ratio of the decay 
$B_{s}\rightarrow\mu^{+}\mu^{-}$ \cite{abazov,tonelli-06}.
The SM prediction for this branching ratio is $(3.2\pm1.5)\times10^{-9}$ 
\cite{buras-03} or $\leq 7.7 \times 10^{-9}$ at $3 \sigma$ level.
$B_{s}\rightarrow\mu^{+}\mu^{-}$ will be one of the important rare B 
decays to be studied by the experiments at the upcoming Large Hadron
Collider (LHC). We expect that the present upper limit will be reduced
significantly in these experiments. 
A non-zero value of this branching ratio is measurable,
if it is $\geq 10^{-8}$ \cite{forty}.

In a previous publication \cite{alok-sankar-05}, we studied the 
constraints on new physics contribution to the branching ratio of 
$B_{s}\rightarrow\mu^{+}\mu^{-}$ coming from the experimentally 
measured values of the branching ratios of 
$B\rightarrow(K,K^{*}) \mu^{+} \mu^{-}$.
We found that if the new physics interactions are in the form of 
vector/axial-vector
operators, then the present data on $B(B\rightarrow(K,K^{*})\mu^{+}\mu^{-})$
does not allow a large boost in $B(B_{s}\rightarrow\mu^{+}\mu^{-})$.
By large boost we mean an enhancement of at least an order of magnitude
in comparison to the SM prediction. However, if the new physics interactions
are in the form of the scalar/pseudoscalar operators, then the presently
measured rates of $B\rightarrow(K,K^{*})\mu^{+}\mu^{-}$ do not put any
useful constraints on $B_s \rightarrow \mu^+ \mu^-$ and 
$B_{NP}(B_{s}\rightarrow\mu^{+}\mu^{-})$ can
be as high as the present experimental upper limit. Therefore we are led to  
the conclusion that if future experiments measure 
$B_{s}\rightarrow\mu^{+}\mu^{-}$ with a branching ratio greater
than $10^{-8}$, then the new physics giving
rise to this decay has to be in the form of scalar/pseudoscalar interaction.

Recently $B_{s}-\bar{B}_{s}$ mixing has been observed experimentally
\cite{Giagu-06}, with a very small experimental error. In this paper, 
we want to  see what constraint this measurement imposes on the new physics 
contribution to the branching ratio of $B_{s}\rightarrow\mu^{+}\mu^{-}$. 
In particular, we consider the question: Does it allow new physics in
the form of scalar/pseudoscalar interaction to give a large boost
in $B_{NP}(B_{s}\rightarrow\mu^{+}\mu^{-})$ ?

We start by considering the $B_{s}\rightarrow\mu^{+}\mu^{-}$ decay.
The effective new physics lagrangian for the quark
level transition $\bar{b}\rightarrow \bar{s}\mu^{+}\mu^{-}$ due to 
scalar/pseudoscalar interactions can arise from tree and/or electroweak
penguin and/or box diagrams. We parametrize it as  
\begin{equation}
L_{\bar{b}\rightarrow \bar{s}\mu^{+}\mu^{-}}^{SP}=G_{1}\,
\bar{b}(g_{S}^{sb}+g_{P}^{sb}
\gamma_{5})s\,\,\bar{\mu}(g_{S}^{\mu\mu}+g_{P}^{\mu\mu}\gamma_{5})\mu,
\label{eq1}
\end{equation}
where $G_1$ is a dimensional factor characterizing the overall scale of
new physics, with dimension $(mass)^{-2}$. 
This factor essentially arises due to the scalar propagator in 
tree or electroweak penguin diagrams (or scalar propagators in box 
diagrams) which couples the quark bilinear to the lepton bilinear. 
$g_{S,P}^{sb}$ and $g_{S,P}^{\mu \mu}$ are dimensionless 
numbers, characterizing, respectively, $b-s$ and $\mu-\mu$ couplings 
due to new physics scalar/pseudoscalar interactions. 
Electromagnetic penguins 
necessarily have vector couplings in the lepton bilinear 
so they do not contribute to the effective lagrangian in eq.~(\ref{eq1}).
The amplitude for the decay $B_{s}\rightarrow l^{+}l^{-}$ is given
by
\begin{equation}
M(B_{s}\rightarrow\mu^{+}\mu^{-})=G_{1}\, g_{P}^{sb}\langle0\left|\bar{b}\gamma_{5}s\right|B_{s}\rangle\left[g_{S}^{\mu\mu}\bar{u}(p_{\mu})v(p_{\bar{\mu}})+g_{P}^{\mu\mu}\bar{u}(p_{\mu})\gamma_{5}v(p_{\overline{\mu}})\right].
\end{equation}
The pseudoscalar matrix element is,
\begin{equation}
\langle0\left|\bar{b}\gamma_{5}s\right|B_{s}\rangle=-i\frac{f_{B_{s}}M_{B_{s}}^{2}}{m_{b}+m_{s}},
\end{equation}
where $m_{b}$ and $m_{s}$ are the masses of bottom and strange quark
respectively. 


The calculation of the decay rate gives
\begin{equation}
\Gamma_{NP}(B_{s}\rightarrow\mu^{+}\mu^{-})=(g_{P}^{sb})^{2}[(g_{S}^{\mu\mu})^{2}+(g_{P}^{\mu\mu})^{2}]\frac{G_{1}^{2}}{8\pi}\frac{f_{B_{s}}^{2}M_{B_{s}}^{5}}{(m_{b}+m_{s})^{2}}.\label{Gamma_Bs}
\end{equation}
We see that the decay rate depends upon the new physics couplings
$(g_{P}^{sb})^{2}$ and $G_{1}^{2}[(g_{S}^{\mu\mu})^{2}+(g_{P}^{\mu\mu})^{2}]$.
To obtain information on these parameters, we look at $B_{s}-\bar{B_{s}}$
mixing together with $K_{L}\rightarrow\mu^{+}\mu^{-}$ decay and
$K^{0}-\bar{K^{0}}$ mixing. 

Let us consider $B_{s}-\bar{B_{s}}$ mixing to obtain a constraint
on $(g_{P}^{sb})^{2}$. Replacing leptonic bilinear by quark bilinear
in eq.~\ref{eq1}, we get $\Delta B=2$ Lagrangian,
\begin{equation}
L_{B_{s}-\bar{B_{s}}}^{SP}=G_{2}\,\bar{b}(g_{S}^{sb}+g_{P}^{sb}\gamma_{5})s\,\,\bar{b}(g_{S}^{sb}+g_{P}^{sb}\gamma_{5})s,
\label{eq5}
\end{equation}
where $G_2$ is another dimensional factor. As in the case of $G_1$, 
introduced in eq.~(\ref{eq1}), $G_2$ also arises due to the scalar 
propagator (or progators in the case of box diagrams). Therefore
it also has dimension $(mass)^{-2}$ and {\it is of the same order of magnitude
as $G_1$.}
From eq.~(\ref{eq5}), we calculate the mass difference of the 
$B_s$ mesons to be 
\begin{equation}
\Delta m_{B_{s}}=\frac{1}{2 M_{B_S}} G_{2}\,(g_{P}^{sb})^{2}\hat{B}_{B_{s}}\frac{f_{B_{s}}^{2}M_{B_{s}}^{4}}{(m_{b}+m_{s})^{2}}.
\end{equation}
Thus the effective $b-s$ pseudoscalar coupling is obtained to be  
\begin{equation}
(g_{P}^{sb})^{2}=\frac{\Delta m_{B_{s}}(m_{b}+m_{s})^{2}}{2\hat{B}_{B_{s}}
f_{B_{s}}^{2}M_{B_{s}}^{3}G_{2}}.
\label{sbps}
\end{equation}

We now consider the decay $K_{L}\rightarrow\mu^{+}\mu^{-}$. The same
new physics leading to the effective $\bar{b}\rightarrow \bar{s}\mu^{+}\mu^{-}$ 
lagrangian in eq.~(\ref{eq1}), also leads a similar effective lagrangian
for $\bar{s}\rightarrow \bar{d}\mu^{+}\mu^{-}$ transition. 
The only difference will be the
effective scalar/pseudoscalar couplings in the quark bilinear. 
Thus we have,
\begin{equation}
L_{\bar{s}\rightarrow \bar{d} \mu^{+}\mu^{-}}^{SP}=G_{1}\,\bar{s}(g_{S}^{sd}+g_{P}^{sd}\gamma_{5})d\,\bar{\mu}(g_{S}^{\mu\mu}+g_{P}^{\mu\mu}\gamma_{5})\mu.
\end{equation}
The calculation of decay rate gives
\begin{equation}
\Gamma_{NP}(K_{L}\rightarrow\mu^{+}\mu^{-})=2(g_{P}^{sd})^{2}[(g_{S}^{\mu\mu})^{2}+(g_{P}^{\mu\mu})^{2}]\frac{G_{1}^{2}}{8\pi}\frac{f_{K}^{2}M_{K}^{5}}{(m_{d}+m_{s})^{2}}.
\label{DR_Kmu}
\end{equation}
Here extra factor of 2 occurs because the amplitudes $A(K^{0}\rightarrow\mu^{+}\mu^{-})=A(\bar{K^{0}}\rightarrow\mu^{+}\mu^{-})$ and $K_{L}=\frac{K^{0}+\bar{K^{0}}}{\sqrt{2}}$.
We see that $G_{1}^{2}[(g_{S}^{\mu\mu})^{2}+(g_{P}^{\mu\mu})^{2}]$ can be
calculated from $\Gamma (K_L \rightarrow \mu^+ \mu^-)$, once we know the 
value of $(g_{P}^{sd})^{2}$. In order
to determine the value of $(g_{P}^{sd})^{2}$, we consider $K^{0}-\bar{K^{0}}$
mixing. The effective scalar/pseudoscalar new physics lagrangian for 
this process can
be obtained from that of $\bar{s}\rightarrow \bar{d}\mu^{+}\mu^{-}$ by replacing
lepton current by corresponding quark current or equaivalently from 
effective lagrangian of eq.~(\ref{eq5}) where $b-s$ quark bilinear 
is replaced by $s-d$ quark bilinear,
\begin{equation}
L_{K^{0}-\bar{K^{0}}}^{SP}=G_{2}\,\bar{s}(g_{S}^{sd}+g_{P}^{sd}\gamma_{5})d\,\,\bar{s}(g_{S}^{sd}+g_{P}^{sd}\gamma_{5})d.
\end{equation}
From this lagrangian, we obtain the $K_L - K_S$ mass difference to be
\begin{equation}
\Delta m_K=\frac{1}{2 M_K} G_{2}\,(g_{P}^{ds})^{2}\hat{B}_K\frac{f_K^{2}M_K^{4}}{(m_{s}+m_{d})^{2}}.
\end{equation}
Thus the effective $s-d$ pseudoscalar coupling is  
\begin{equation}
(g_{P}^{sd})^{2}=\frac{2\Delta m_{K}(m_{d}+m_{s})^{2}}{\hat{B_{K}}f_{_{K}}^{2}M_{K}^{3}G_{2}}.
\end{equation}
Substituting the above value of $(g_{P}^{sd})^{2}$ in eq.~(\ref{DR_Kmu}),
we get
\begin{equation}
G_{1}^{2}[(g_{S}^{\mu\mu})^{2}+(g_{P}^{\mu\mu})^{2}]=
\frac{2\pi G_{2}\hat{B_{K}}}{M_{K}^{2}\Delta m_{K}}\Gamma_{NP}(K_{L}\rightarrow\mu^{+}\mu^{-}).
\label{g1sq}
\end{equation}
Substituting the value of $G_{1}^{2}[(g_{S}^{\mu\mu})^{2}+(g_{P}^{\mu\mu})^{2}]$
from eq.~(\ref{g1sq}) and $(g_{P}^{sb})^{2}$ from eq.~(\ref{sbps}) in eq.~(\ref{Gamma_Bs}), we get
\begin{equation}
\Gamma_{NP}(B_{s}\rightarrow\mu^{+}\mu^{-})={\frac{1}{2}\left(\frac{M_{B_{s}}}{M_{K}}\right)}^{2}\left(\frac{\Delta m_{B_{s}}}{\Delta m_{K}}\right)\left(\frac{\hat{B_{K}}}{\hat{B}_{B_{s}}}\right)\Gamma_{NP}(K_{L}\rightarrow\mu^{+}\mu^{-}).
\end{equation}
The branching ratio is given by,
\begin{equation}
B_{NP}(B_{s}\rightarrow\mu^{+}\mu^{-})={\frac{1}{2}\left(\frac{M_{B_{s}}}{M_{K}}\right)}^{2}\left(\frac{\Delta m_{B_{s}}}{\Delta m_{K}}\right)\left(\frac{\hat{B_{K}}}{\hat{B}_{B_{s}}}\right)
\left[\frac{\tau(B_{s})}{\tau(K_{L})}\right]
B_{NP}(K_{L}\rightarrow\mu^{+}\mu^{-}).
\label{bnp}
\end{equation}

We wish to obtain the largest possible value for $B(B_s \rightarrow 
\mu^+ \mu^-)$. To this end, we make the 
liberal assumption that the experimental values of $\Delta m_{B_{s}}$, 
$\Delta m_{K}$ and $B_{NP}(K_{L}\rightarrow\mu^{+}\mu^{-})$ are saturated 
by new physics couplings. The decay rate for $K_L \rightarrow \mu^+ \mu^-$
consists of both long distance and short distance contributions. The new
physics we consider here, contributes only to the short distance part 
of the decay rate. In ref \cite{isidori-03}, an upper limit on the short 
distance contribution to $B(K_L \rightarrow \mu^+ \mu^-)$ is calculated 
to be $2.5\times10^{-9}$. The mass difference of the $B_s$ mesons is 
recenly measured by the CDF collaboration to be  
$\Delta m_{B_{s}}=(1.17\pm0.01)\times10^{-11}\, GeV$
\cite{Giagu-06}. The bag parameters for the $K$ and the $B_s$ mesons
are $\hat{B_{K}}=(0.58\pm0.04)$ and $\hat{B}_{B_{s}}=(1.30\pm0.10)$
\cite{hashimoto}. The values of the other parameters
of eq.~(\ref{bnp}) are taken from Review of Particle Properties \cite{pdg}:
$\Delta m_{K}=(3.48\pm0.01)\times10^{-15}\, GeV$
$\tau(B_{s})=(1.47\pm0.06)\times10^{-12}\, Sec$
and $\tau(K_{L})=(5.11\pm0.02)\times10^{-8}\, Sec$. 
Substituting these values in eq.~(\ref{bnp}), we get 
\begin{equation}
B_{NP}(B_{s}\rightarrow\mu^{+}\mu^{-})=(6.30\pm0.75)\times10^{-9},
\end{equation}
where all the errors are added in quadrature.
At $3\sigma$, $B_{SM}(B_{s}\rightarrow\mu^{+}\mu^{-})<7.7\times10^{-9}$
where as $B_{NP}(B_{s}\rightarrow\mu^{+}\mu^{-})<8.55\times10^{-9}$. 
Thus we see that this upper bound is almost the same as the SM prediction
even if we maximize the new physics couplings by assuming that they 
saturate the experimental values.
Therefore the present data on $B_{s}-\bar{B}_{s}$ mixing together with data on 
$K^{0}-\bar{K^{0}}$ mixing and $K_{L}\rightarrow\mu^{+}\mu^{-}$
decay puts a strong constraint on new physics scalar/pseudoscalar couplings
and doesn't allow a large boost in the branching ratio of 
$B_{s}\rightarrow\mu^{+}\mu^{-}$.

We now assume that the new physics involving scalar/pseudoscalar couplings
accounts for the difference between the experimental values and the SM
predictions of $\Delta m_K$, $\Delta m_{B_s}$ and the short distance 
contribution to $\Gamma (K_L \rightarrow \mu^+ \mu^-)$. The SM value for 
$B_{s}-\bar{B_{s}}$ is given by 
\cite{monika-06,buras-90},
\begin{equation}
(\Delta m_{B_{s}})_{SM}=\frac{G_{F}^{2}}{6\pi^{2}}\eta_{B}M_{B_{s}}\left(\hat{B}_{B_{s}}f_{B_{s}}^{2}\right)M_{W}^{2}S(x_{t})\left|V_{ts}\right|^{2}\,=\,(1.16\pm0.32)\times10^{-11}\, GeV,
\end{equation}
with $f_{B_{s}}\sqrt{\hat{B}_{B_{s}}}=(262\pm35)\, MeV$ \cite{hashimoto}
, $\eta_{B}=0.55\pm0.01$\cite{buras-90} and 
$\left|V_{ts}\right|=0.0409\pm0.0009$ \cite{pdg}. 
$S(x_{t})$ with $x_{t}=m_{t}^{2}/m_{W}^{2}$ is one of the Inami-Lim
functions \cite{inami-lim}. 
The SM value for $K^{0}-\bar{K^{0}}$ mixing is 
given by \cite{buras-05},
\begin{equation}
(\Delta m_{K})_{SM}=\frac{G_{F}^{2}}{6\pi^{2}}\left(\hat{B}_{K}f_{K}^{2}\right)M_{K}M_{W}^{2}\left[\lambda_{c}^{*2}\eta_{1}S(x_{c})+\lambda_{t}^{*2}\eta_{2}S(x_{t})+2{\lambda_{c}^{*}\lambda}_{t}^{*}\eta_{3}S(x_{c},x_{t})\right],
\end{equation}
where $\lambda_{j}=V_{js}^{*}V_{jd}$, $x_{j}=m_{j}^{2}/m_{W}^{2}$.
The functions $S$ are given by \cite{buras84,buras_84},
\begin{equation}
S(x_{t})=2.46\left(\frac{m_{t}}{170\, GeV}\right)^2,\,\,\,\, S(x_{c})=x_{c}.
\end{equation}
\begin{equation}
S(x_{c},x_{t})=x_{c}\left[\ln\frac{x_{t}}{x_{c}}-\frac{3x_{t}}{4(1-x_{t})}-\frac{3x_{t}^{2}\ln x_{t}}{4(1-x_{t})^{2}}\right].
\end{equation}
Using $\eta_{1}=(1.32\pm0.32)$ \cite{herrlich-94}, $\eta_{2}=(0.57\pm0.01)$
\cite{buras-90}, $\eta_{3}=(0.47\pm0.05)$ \cite{herrlich-95,herrlich-96},
$\hat{B_{K}}=(0.58\pm0.04)$ \cite{hashimoto} ; $f_{K}=(159.8\pm1.5)\, MeV$,
$\left|V_{cs}\right|=0.957\pm0.017\pm0.093$, 
$\left|V_{cd}\right|=0.230\pm0.011$,
$\left|V_{ts}\right|=0.0409\pm0.0009$ and 
$\left|V_{td}\right|=0.0074\pm0.0008$ \cite{pdg}, we get
\begin{equation}
(\Delta m_{K})_{SM}=(1.87\pm0.49)\times10^{-15}\, GeV.
\end{equation}
All the masses were taken from \cite{pdg}.
Considering only the short-distance effects, the SM branching ratio
for $K_{L}\rightarrow\mu^{+}\mu^{-}$ in next-to-next-to-leading order
of QCD is $(0.79\pm0.12)\times10^{-9}$ \cite{gorbahn-06}. Substracting
out the SM contribution from the experimental values of $\Delta m_{B_{s}}$,
$\Delta m_{K}$ and $B_{NP}(K_{L}\rightarrow\mu^{+}\mu^{-})$ , we
get
\begin{eqnarray}
B_{NP}(B_{s}\rightarrow\mu^{+}\mu^{-}) & = & 
\left[ \frac{(\Delta m_{B_s})_{exp} - (\Delta m_{B_s})_{SM}}{
(\Delta m_K)_{exp} - (\Delta m_K)_{SM}} \right]
{\frac{1}{2}\left(\frac{M_{B_{s}}}{M_{K}}\right)}^{2}
\left(\frac{\hat{B_{K}}}{\hat{B}_{B_{s}}}\right)
\left[\frac{\tau(B_{s})}{\tau(K_{L})}\right] \nonumber \\
& & 
\left(B_{exp}(K_{L}\rightarrow\mu^{+}\mu^{-})_{short} -
B_{SM}(K_{L}\rightarrow\mu^{+}\mu^{-})\right).
\end{eqnarray}
Substituting the experimental values and the SM predictions in the
above equation, and adding all the errors in quadrature, we get 
\begin{equation}
B_{NP}(B_{s}\rightarrow\mu^{+}\mu^{-})=(0.08\pm2.54)\times10^{-9}.
\end{equation}
which is consistent with zero. At $3 \sigma$, the upper limit on the
new physics contribution is close to 
SM prediction. Thus the present data on 
$\Delta m_{B_{s}}$ along with $\Delta m_{K}$ and 
$B_{NP}(K_{L}\rightarrow\mu^{+}\mu^{-})$ puts strong constraints on 
new physics scalar/pseudoscalar couplings and doesn't allow a large 
enhancement in the branching ratio of $B_{NP}(B_{s}\rightarrow\mu^{+}\mu^{-})$
much beyond the SM predictions. \textbf{\emph{New physics at most
can cause a factor of two enhancement but not an order of magnitude}}. Hence 
the total branching ratio which is the sum of SM contribution and new physics
contribution will be of the order of $10^{-8}$ and hence reachable at LHC.

\emph{Conclusions:}

In this letter, we considered the constraints on the New Physics
couplings of scalar/pseudoscalar type in the $b \rightarrow s$ 
transition. It was shown previously that only such New Physics
can give rise to an order of magnitude enhancement of the decay
rate for $B_s \rightarrow \mu^+ \mu^-$. Using the recent data on 
$B_s - \bar{B}_s$ mixing, together with the data on $K^0 - \bar{K}^0$ 
mixing and the short distance contribution to $K_L \rightarrow \mu^+ 
\mu^-)$, we obtained very strong bounds on $B( B_s \rightarrow \mu^+ 
\mu^-)$. New Physics in the form of scalar/pseudoscalar couplings 
can at most increase the $B(B_s \rightarrow \mu^+ \mu^-)$ by a 
factor of $2$ compared to its Standard Model prediction. An order
magnitude enhancement, previously allowed, is ruled out. 

\begin{acknowledgments}
We thank Prof. Rohini Godbole for posing a 
question which led to this investigation. We also thank Prof.
B. Ananthanarayan for a critical reading of the manuscript. 
\end{acknowledgments}

\end{document}